\documentclass[prd,aps,showpacs,preprintnumbers,amssymb]{revtex4}

\usepackage{epsf}

\def\e3p{$\eta \rightarrow 3 \pi$}

\begin{document}

\title{%
\hfill{\normalsize\vbox{%
\hbox{\rm SU-4252-858}
}}\\
{
Low energy scattering with a nontrivial pion
}}

\author{Amir H. Fariborz $^{\it \bf a}$~\footnote[3]{Email:
fariboa@sunyit.edu}}

\author{Renata Jora $^{\it \bf b}$~\footnote[2]{Email:
cjora@physics.syr.edu}}

\author{Joseph Schechter $^{\it \bf c}$~\footnote[4]{Email:
schechte@physics.syr.edu}}

\affiliation{$^ {\bf \it a}$ Department of Mathematics/Science,
State University of New York Institute of Technology, Utica,
NY 13504-3050, USA.}

\affiliation{$^ {\bf \it b,c}$ Department of Physics,
Syracuse University, Syracuse, NY 13244-1130, USA,}

\date{\today}

\begin{abstract}

An earlier calculation in a generalized linear sigma
model showed that the well-known current
algebra formula for low energy pion pion scattering
held even though the massless
Nambu Goldstone pion
contained a small admixture of a
two-quark two-antiquark field.  Here we turn on the pion
mass and note that the current algebra  formula
no longer holds exactly.  We discuss this small deviation
and also study the effects of an SU(3) symmetric quark mass
type
term on the masses and mixings of the eight SU(3) multiplets
in the model. We calculate the s wave scattering
lengths, including the beyond current algebra theorem
corrections due to the scalar mesons,
and observe that the model
can fit the data well. In the process, we uncover the
way in which linear sigma models give controlled
corrections (due to the presence of scalar mesons) to the
current algebra scattering formula. Such a feature
 is commonly thought to exist only in the non-linear
sigma model approach. 
\end{abstract}


\maketitle

\section{Introduction}

    A linear sigma model with both quark-antiquark type
fields and fields containing (in an unspecified configuration)
two quarks and two antiquarks, seems useful for
understanding the light
scalar spectrum of QCD. In a previous treatment
\cite{bigpaper} we considered
a usual simplification in which the three light quark masses
were taken to be zero. The model was seen to give a neat
intuitive explanation of how ``four quark" scalar states could
be naturally much lighter than the conventional p-wave
quark-antiquark scalars. 
We also verified in detail that, as long
as the potential of the model satisfied SU(3)$_{\rm L}$ $\times$
SU(3)$_{\rm R}$ invariance, the massless version of the famous current
algebra theorem \cite{W} on low energy pion pion scattering was correct.

   In the present paper we introduce a common mass for the three
light quarks in such a way that the pion gets its correct mass.
SU(3) flavor invariance continues to hold, which is a desirable
simplification. First we reexamine the masses and mixings of the
particles in the model. It is seen that the natural
explanation for the lightness of a ``four quark"
scalar remains unchanged.
 Then we reexamine the pion pion scattering
amplitude to try to see if
the low energy theorem
continues to hold. Curiously we find that it does not exactly
hold. What goes wrong? The algebra of the Noether currents
should be good in this model so that is not the cause.
It turns out that the partially conserved axial vector
current, which is also required for the theorem,
does not hold, unlike for the massless case.
The axial vector current has a single particle contribution from
the ``heavy pion" in this chiral model as well as from the ordinary
pion. The ordinary pion does not therefore completely saturate
the axial current. Actually this is a small effect but is
of conceptual interest and may be of more importance for the kaon
scattering case.

    A more important quantitative effect arises from the
contributions of the scalar isosinglet mesons in the model.
It is shown that
these contributions can explain the experimental s wave
isosinglet scattering length. 

    The notation is reviewed in Section II.
The corrections to the masses and mixings,
due to the quark mass term, are studied in Section III.
An approximate analytic treatment of the scattering, for
a general potential, is contained in Section IV. The
exact numerical treatment for a ``leading order"
potential is presented in Section 
V.
Some discussion and conclusions are given in Section VI.
The Appendix explains the method of parameter determination
from experiment and a listing of typical values for all
the parameters.

\section{Notation}
    We introduce the 3$\times$3 matrix
chiral nonet fields;
\begin{equation}
M = S +i\phi, \hskip 2cm
M^\prime = S^\prime +i\phi^\prime.
\label{sandphi}
\end{equation}
Here $M$ represents scalar, $S$ and pseudoscalar,
$\phi$ quark-antiquark type states, while
$M^\prime$ represents states which are made of
two quarks and two antiquarks.    The transformation
properties under  SU(3)$_{\rm L}\times$ SU(3)$_{\rm R}
\times$ U(1)$_{\rm A}$ are
\begin{equation}
M \rightarrow e^{2i\nu}
\, U_{\rm L} M U_{\rm R}^\dagger, \hskip 2cm
M^\prime \rightarrow e^{-4i\nu}
\, U_{\rm L} M^\prime U_{\rm R}^\dagger,
\label{Mchiral}
\end{equation}
where $U_{\rm L}$ and $U_{\rm R}$ are unitary unimodular
matrices, and the phase $\nu$ is associated with the
U(1)$_{\rm A}$ transformation.
The general Lagrangian density which defines our model is
\begin{equation}
{\cal L} = - \frac{1}{2} {\rm Tr}
\left( \partial_\mu M \partial_\mu M^\dagger
\right) - \frac{1}{2} {\rm Tr}
\left( \partial_\mu M^\prime \partial_\mu M^{\prime \dagger} \right)
- V_0 \left( M, M^\prime \right) - V_{SB},
\label{mixingLsMLag}
\end{equation}
where $V_0(M,M^\prime) $ stands for a function made
from SU(3)$_{\rm L} \times$ SU(3)$_{\rm R}$
(but not necessarily U(1)$_{\rm A}$) invariants
formed out of
$M$ and $M^\prime$. The quantity $V_{SB}$ stands for
chiral symmetry
breaking terms which transform in the same way
as the quark mass terms in
the fundamental QCD Lagrangian.   In our previous paper
\cite{bigpaper}, we focused on general properties which
continued to hold
when $V_{SB}$ was set to zero.    Here, we
include the SU(3) symmetric mass term:
\begin{equation}
V_{SB} = - 2\, A\, {\rm Tr} (S)
\label{vsb}
\end{equation}
where $A$ is a real parameter.    A characteristic
feature of the model is the presence of ``two-quark'' and
``four-quark'' condensates:
\begin{equation}
\left\langle S_a^b \right\rangle = \alpha_a \delta_a^b,
\quad \quad \left\langle S_a^{\prime b} \right\rangle =
\beta_a \delta_a^b.
\label{vevs}
\end{equation}
We shall assume the vacuum to be SU(3)$_{\rm V}$
invariant, which implies
\begin{equation}
\alpha_1 = \alpha_2 = \alpha_3 \equiv \alpha, \hskip 2cm
\beta_1 = \beta_2 = \beta_3 \equiv \beta.
\end{equation}
The SU(3) particle content of the model consists of two
pseudoscalar octets, two pseudoscalar singlets, two scalar
octets  and two scalar singlets.   This gives us eight
different masses and four mixing angles.    We next give
the notations for resolving the nonets into SU(3) octets
and singlets.    Note the matrix convention $\phi_a^b \to
\phi_{ab}$.    The properly normalized singlet states
are:
\begin{eqnarray}
\phi_0=\frac{1}{\sqrt{3}}{\rm Tr}(\phi),\hspace{1cm}
\phi_0^\prime=\frac{1}{\sqrt{3}}{\rm Tr} (\phi^\prime),
\nonumber \\
S_0=\frac{1}{\sqrt{3}}{\rm Tr}(S),\hspace{1cm}
S_0^\prime=\frac{1}{\sqrt{3}}{\rm Tr}(S^\prime).
\label{singlets}
\end{eqnarray}
Then we have the matrix decompositions:
\begin{eqnarray}
\phi={\hat \phi}+\frac{1}{\sqrt{3}}\phi_{0}1,\hspace{1cm}
\phi^\prime={\hat \phi^\prime}+\frac{1}{\sqrt{3}}\phi_{0}^\prime1,
\nonumber \\
S={\hat S}+\frac{1}{\sqrt{3}}S_{0}1,\hspace{1cm}
S^\prime={\hat S^\prime}+\frac{1}{\sqrt{3}}S_{0}^{\prime}1,
\label{octets}
\end{eqnarray}
wherein ${\hat \phi}$, ${\hat \phi^\prime}$, ${\hat S}$ and
${\hat S^\prime}$
are all 3 $\times $ 3 traceless matrices.
The singlet scalar fields may be further
decomposed as:
\begin{equation}
S_0=\sqrt{3}\alpha+{\tilde S_0},\hspace{1cm}
S_0^\prime=\sqrt{3}\beta+{\tilde S_0}^\prime.
\label{scalarsinglets}
\end{equation}
Here ${\tilde S_0}$ and ${\tilde S_0}^\prime$ are the
fluctuation fields around the true ground state of the model.
The breaking of SU(3) to the isospin group SU(2) will be
examined in the future.     In that case there are 16
different masses, four 2$\times$2 mixing matrices and two
4$\times$4 mixing matrices.
To fully characterize the system we will also require some
knowledge of the axial vector and vector currents obtained by
Noether's method:
\begin{eqnarray}
(J_\mu^{axial})_a^b &=&(\alpha_a+\alpha_b)\partial_\mu\phi_a^b +
(\beta_a+\beta_b)\partial_\mu{\phi'}_a^b+ \cdots,
\nonumber \\
(J_\mu^{vector})_a^b &=&i(\alpha_a-\alpha_b){\partial_\mu} S_a^b +
i(\beta_a-\beta_b)\partial_\mu {S'}_a^b+ \cdots,
\label{currents}
\end{eqnarray}
where the dots stand for terms bilinear in the fields.

    In our model we use a previously discussed scheme to select the most
important terms in the potential, $V_0(M,M')$. The favored terms which
are SU(3)$_{\rm L}\times$SU(3)$_{\rm R}$ invariant but violate U(1)$_{\rm
A}$ are:
\begin{equation}
V_\eta=c_3\, [F_\eta(M,M')]^2,
\label{veta}
\end{equation}
in which $c_3$ is a coupling constant and
\begin{equation}
F_\eta(M,M')=
\gamma_1\,   
 {\rm ln} \left(
      {
         { {\rm det}(M)}
             \over
        { {\rm det}(M^\dagger)}
      }
\right)
+(1-\gamma_1) \,
{\rm ln}\left(
 {
   { {\rm Tr}(MM'^\dagger) }
            \over
   { {\rm Tr}(M'M^\dagger) }
 }
\right),
\label{gamma1}
\end{equation}
where $\gamma_1$ is a dimensionless parameter.
This form exactly mocks up the U(1)$_{\rm
A}$ anomaly of QCD.
Information about the pseudoscalar particles which is independent
of the choice of the  U(1)$_{\rm A}$ invariant terms in $V_0$ may be
obtained by
differentiating
the following matrix equation representing the response of the potential
to an infinitesimal axial transformation:
\begin{equation}
\left[\phi,\frac{\partial V_0}{\partial S}\right]_+ -
\left[S,\frac{\partial V_0}{\partial \phi}\right]_+ +
(\phi,S)\rightarrow(\phi',S') =
1\left[ 2\, {\rm Tr} \left(\phi'\frac{\partial V_0}{\partial S'}-
S'\frac{\partial V_0}{\partial \phi'}\right) - 8\,c_3\, i\, 
F_\eta(M,M')\right].
\label{geneq}
\end{equation}
To get general constraints on the pseudoscalar particle masses we
differentiate
this equation once with respect to each of the two matrix fields:
$\phi,\phi'$ and evaluate the equation in the ground state.
Thus we also need the ``minimum" condition,
\begin{equation}
\left< \frac{\partial V_0}{\partial S}\right> + \left< \frac{\partial 
V_{SB}}{\partial
S}\right>=0,
\quad \quad \left< \frac{\partial V_0}{\partial S'}\right> + 
\left<\frac{\partial
V_{SB}}{\partial S'}\right> =0.
\label{mincond}
\end{equation}

\section{Masses and mixings}
    As we previously discussed, the leading choice of terms corresponding
to eight or fewer quark plus antiquark lines at each effective vertex
reads:
\begin{eqnarray}
V_0 =&-&c_2 \, {\rm Tr} (MM^{\dagger}) +
c_4^a \, {\rm Tr} (MM^{\dagger}MM^{\dagger})
\nonumber \\
&+& d_2 \,
{\rm Tr} (M^{\prime}M^{\prime\dagger})
     + e_3^a(\epsilon_{abc}\epsilon^{def}M^a_dM^b_eM'^c_f + h.c.)
\nonumber \\
     &+&  c_3\left[ \gamma_1 {\rm ln} (\frac{{\rm det} M}{{\rm det}
M^{\dagger}})
+(1-\gamma_1)\frac{{\rm Tr}(MM'^\dagger)}{{\rm Tr}(M'M^\dagger)}\right]^2.
\label{SpecLag}
\end{eqnarray}
     All the terms except the last two have been chosen to also
possess the  U(1)$_{\rm A}$
invariance.

The minimum equations for this potential are:
\begin{equation}
\left\langle { {\partial V_0} \over {\partial S_a^a} } \right\rangle =
2 \,\alpha\,  \left( - c_2 + 2\, c_4^a\, \alpha^2 + 4\, e_3^a \, \beta
\right) = 2A,
\label{mealpha}
\end{equation}

\begin{equation}
\left\langle { {\partial V_0} \over
{\partial {S'}_a^a} } \right\rangle
=
2  \left( d_2\, \beta + 2\, e_3^a\, \alpha^2 \right) = 0.
\label{mebeta}
\end{equation}

    Differentiating the potential in Eq.(\ref{SpecLag}) twice
will yield four 2$\times$2 mass matrices denoted as
$(M_\pi^2)$, $(M_0^2)$, $(X_a^2)$ and $(X_0^2)$ respectively
for the pseudoscalar octets, the pseudoscalar singlets,
the scalar octets and the scalar singlets. These may be brought to
diagonal (hatted) form by the following 2$\times$2 orthogonal
transformations:
\begin{eqnarray}
\sum_{B,C}(R_{\pi}^{-1})_{AB}(M_{\pi}^2)_{BC}(R_{\pi})_{CD}&=&({\hat
M}_{\pi}^2)_{AD},\hskip 0.5cm
\sum_{B,C}(R_0^{-1})_{AB}(M_0^2)_{BC}(R_0)_{CD}=({\hat
M}_0^2)_{AD},
\nonumber \\
\sum_{B,C}(L_a^{-1})_{AB}(X_a^2)_{BC}(L_a)_{CD}&=&({\hat
X}_a^2)_{AD}, \hskip 0.5cm
\sum_{B,C}(L_0^{-1})_{AB}(X_0^2)_{BC}(L_0)_{CD}=({\hat
X}_0^2)_{AD},
\label{simtransf}
\end{eqnarray}
Notice that the four mass matrices are identical to those given in section
IV of \cite{bigpaper} for the zero quark mass case. The numerical values 
of
the entries will however
differ because the relations among the coefficients are different due to
the presence of 2A rather than zero on the right hand side of
Eq.(\ref{mealpha}).

  In the massive case there are 9 parameters (A, $\alpha$, $\beta$, $c_2$,
$d_2$,
$c_4^a$, $e_3^a$, $c_3$ and $\gamma_1$). These can be reduced to seven
by use of the two minimum equations just given. We note that
the parameters  $c_3$ and $\gamma_1$, associated with modeling the
U(1)$_{\rm A}$
anomaly, do not contribute to either the minimum equations or to the
mass matrices of the particles which are not $0^-$ singlets. Thus it is
convenient to first determine the other five independent parameters.
As the corresponding experimental inputs
\cite{ropp} we take the non-strange
quantities:
\begin{eqnarray}
m(0^+ {\rm octet}) &=& m[a_0(980)] = 984.7 \pm 1.2\, {\rm MeV}
\nonumber \\
m(0^+ {\rm octet}') &=& m[a_0(1450)] = 1474 \pm 19\, {\rm MeV}
\nonumber \\
m(0^- {\rm octet}') &=& m[\pi(1300)] = 1300 \pm 100\, {\rm MeV}
\nonumber \\
m(0^- {\rm octet}) &=& m_\pi = 137 \, {\rm MeV}
\nonumber \\
F_\pi &=& 131 \, {\rm MeV}
\label{inputs1}
\end{eqnarray}

Evidently, a large experimental uncertainty
appears in the mass of $\pi(1300)$; we shall initially take the other
masses as fixed at their central values and vary this mass in the
indicated range. Essentially $m[\pi(1300)]$ is being
treated as an arbitrary parameter of our model.
As shown in Eq.(\ref{lagpara}) in Appendix A, it is
straightforward to determine the five independent parameters
in terms of these five inputs. This determination is a
generalization of
the one in the previous zero mass pion case in which
four parameters were determined from four inputs.

\begin{figure}[t]
\begin{center}
\vskip 1cm
\epsfxsize = 7.5cm
\ \epsfbox{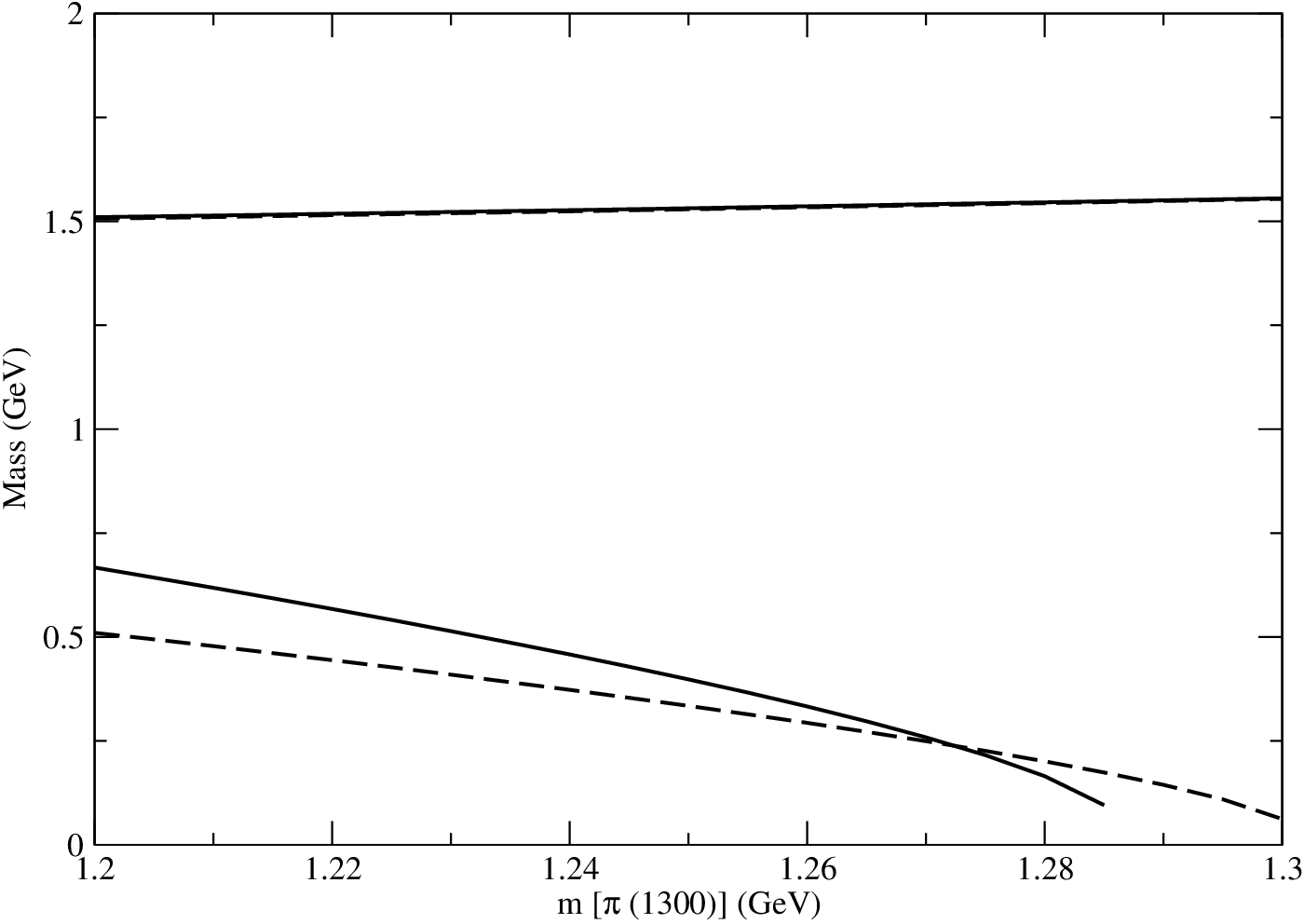}
\end{center}
\caption[]{%
The predictions for the masses of the
two SU(3) singlet scalars
vs. $m[\pi(1300)]$. The solid lines correspond to the massive
pion case while the dashed lines correspond to
the massless pion case previously considered.
}
\label{ms0vsmpip}
\end{figure}

The effects of adding a non zero quark mass term
on the masses of the two predicted scalar singlets are
displayed in Fig.\ref{ms0vsmpip}. It is clear that the small
mass term has a negligible effect on the mass of the
heavier scalar singlet. On the other hand, there
is a larger effect
on the mass of the lighter scalar singlet. Still this
singlet is exceptionally light so there is no
qualitative difference in the result.

\begin{figure}[h]
\begin{center}
\epsfxsize = 7.5cm
\ \epsfbox{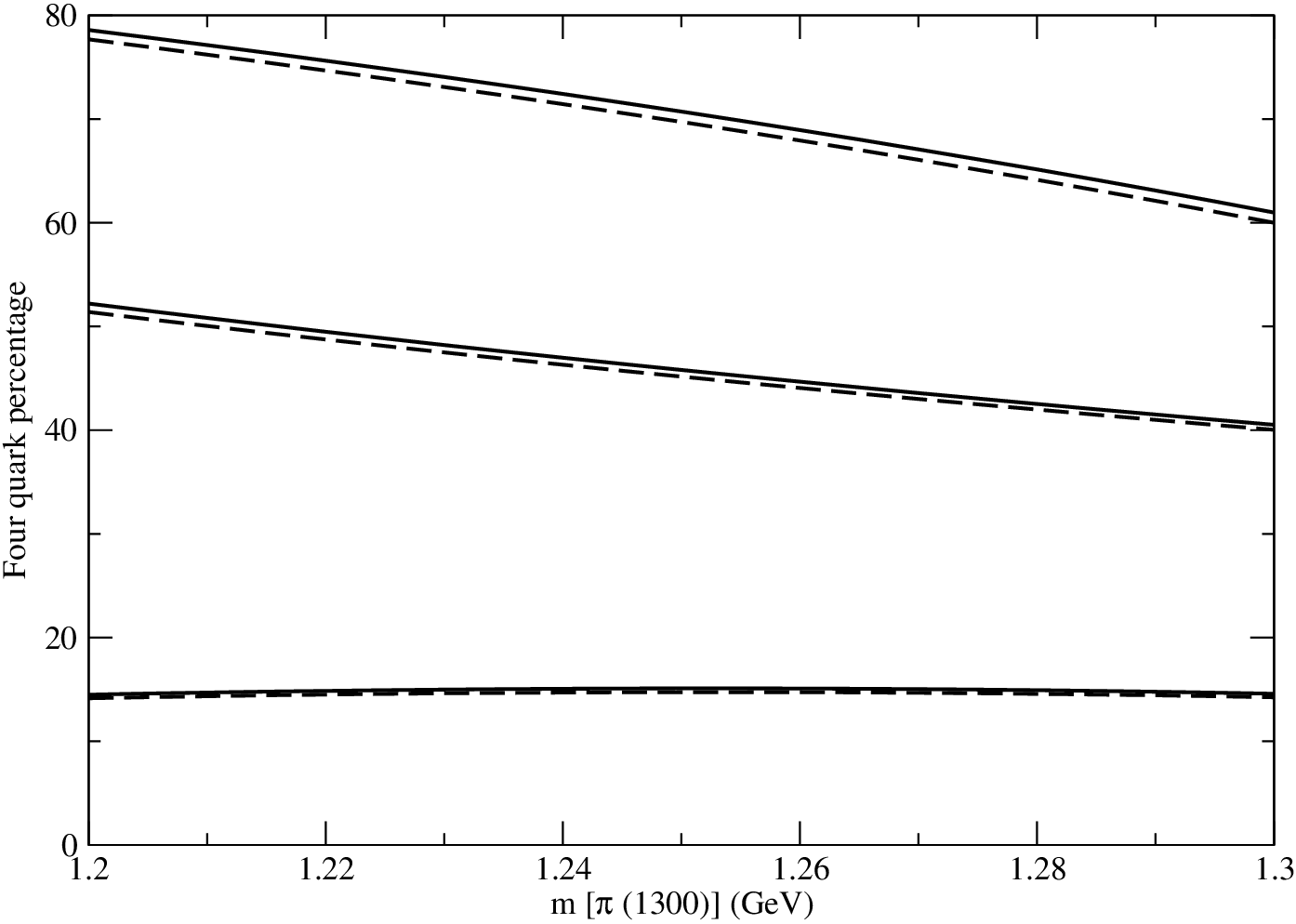}
\end{center}
\caption[]{%
Plot of the four quark percentages of various particles in the model
as functions of the undetermined input parameter, $m[\pi(1300)]$.
Starting from the bottom and going up, the curves respectively
show the four quark percentages of the pion, the
lighter $0^+$ singlet,
and the $a_0(980)$. Solid lines apply for the massive pion case
and dashed lines for the massless pion case.
}
\label{4qp}
\end{figure}

    In Fig. \ref{4qp} we display
a comparison of the four quark percentages of the $\pi$
meson, the lighter $a_0$ meson and the lighter
scalar singlet with the corresponding values in the
model with zero quark masses.
(These are, of course, equal to the two quark percentages
of the heavier particles with the same quantum numbers).
It is clear that there is not much change compared to the
zero quark mass case.
 
\vspace{1cm}

\begin{figure}[h]
\begin{center}
\vskip 1cm
\epsfxsize = 7.5cm
\ \epsfbox{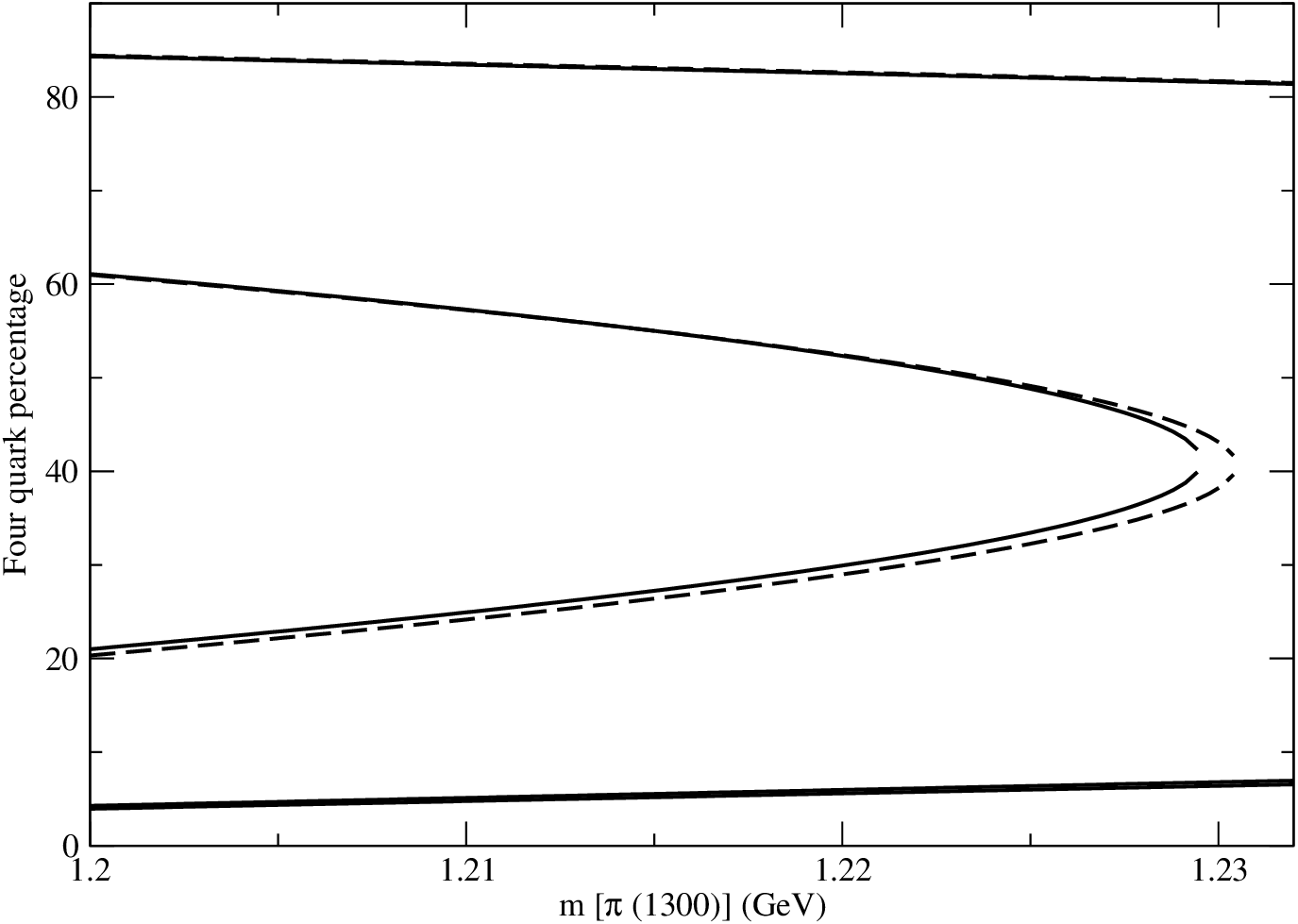}
\end{center}
\caption[]{%
Plot of the four quark percentages of
the $\eta(958)$
as functions of the undetermined input parameter, $m[\pi(1300)]$
for two scenarios. The top and bottom curves correspond to choosing
the $\eta(1760)$ as the heavier $0^-$ SU(3) singlet while the middle two
curves correspond
to choosing the $\eta(1475)$ as the heavier $0^-$ SU(3) singlet.
Note that for each scenario, the two curves are associated with different
solutions of the quadratic equation (\ref{findgamma1}) for $\gamma_1$.
The solid lines apply to the massive pion case
while the dashed lines apply to the previous massless pion case. }
\label{4qpeta}
\end{figure}

It remains to discuss the
four quark percentages of the two SU(3) singlet pseudoscalars.
The lightest is the $\eta$(958) while candidates for the heavier
one include $\eta$(1295), $\eta$(1405), $\eta$(1475) and
$\eta$(1760). As in the zero mass case, the first two candidates
are ruled out because they do not lead to positive eigenvalues
of the prediagonal squared mass matrix $(M_0^2)$. For the other
two scenarios we may find the numerical values of the remaining
parameters $c_3$ and $\gamma_1$ by using Eqs. (\ref{phizeromixing}),
(\ref{K}),
(\ref{findgamma1})
and (\ref{findc3})
given
in the Appendix. The four quark contents for the $\eta$(958)
are being compared between the massive and massless
quark cases in Fig. \ref{4qpeta}. Note that there
are two solutions for each scenario corresponding to
Eq. (\ref{findgamma1}) being of quadratic type. The lower
four quark percentage curves seem the most plausible.
Again there seems to be little difference between
the zero
and non-zero quark mass cases. This is understandable by
comparing the values of the Lagrangian parameters found in
Appendix A with those found in Appendix B of \cite{bigpaper}.

\section{Pion scattering: approximate analytic treatment
for general V$_0$}

    We have seen that there are not very big changes
when using an SU(3) symmetric quark mass term of proper strength to
give the experimental mass value to the pion. However, for the
discussion of the pion pion scattering amplitude near threshold,
which is one of the most important applications of the chiral
approach, the correct value of the pion mass is important.
In the zero quark mass case we showed that for any choice of
terms in the potential V$_0$, the current algebra expression for
the threshold pion pion amplitude held exactly. This was
understandable since the algebra of chiral currents held by
construction and furthermore the pion completely saturated
the axial current. In the present case, the pion does not, as
we shall see, completely saturate the axial current. Thus it is
interesting to study this case in more detail.

  Note that the transformation between the diagonal
fields ($\pi^+$ and $\pi'^+$)  and the
original pion fields is given as:
\begin{equation}
\left[
\begin{array}{c}  \pi^+ \\
                 \pi'^+
\end{array}
\right]
=
R_\pi^{-1}
\left[
\begin{array}{c}
                        \phi_1^2 \\
                        {\phi'}_1^2
\end{array}
\right]=
\left[
\begin{array}{c c}
                \cos\theta_\pi & -\sin \theta_\pi
\nonumber               \\
\sin \theta_\pi & \cos \theta_\pi
\end{array}
\right]
\left[
\begin{array}{c}
                        \phi_1^2 \\
                        {\phi'}_1^2
\end{array}
\right].
\label{mixingangle}
\end{equation}
The value of the mixing angle, $\theta_\pi$ was
previously written for an arbitrary
potential V$_0$ and with the symmetry breaker,
Eq.(\ref{vsb}) as:
\begin{equation}
\tan(2\theta_\pi)=\frac{-2y_\pi z_\pi}{y_\pi(1-z_\pi^2)-x_\pi},
\label{thetasubpi}
\end{equation}
wherein,
\begin{equation}
x_\pi = \frac{2A}{\alpha},\hskip .5cm
y_\pi =\left< \frac{\partial^2V}{\partial
{\phi'}_2^1\partial{\phi'}_1^2}
\right>, \hskip .5cm
z_\pi= \frac{\beta}{\alpha}.
\label{xyzpi}
\end{equation}
The specific values of $x_\pi$,  $y_\pi$
and  $z_\pi$ will depend on the particular
potential. In the case of no symmetry breaking,
there was a big simplification and tan$\theta_\pi$
was given by just  -$\beta/\alpha$. Furthermore, $x_\pi$
and $y_\pi$
would be respectively the squared pion mass
and the squared $\pi(1300)$ mass
in the absence of mixing. Clearly, the ratio
$x_\pi/y_\pi$ is a very small number in the
general case. We now
make use of this fact to solve for the first
correction to the mixing angle obtained
from  Eq.(\ref{thetasubpi}):
\begin{eqnarray}
\sin\theta_\pi&=&\frac{-\beta}{\sqrt{\alpha^2+\beta^2}}
\left[
1+\frac{\alpha^4}{\left(\alpha^2+\beta^2\right)^2}\frac{x_\pi}{y_\pi}
\right]
+\cdots,
\nonumber  \\
\cos\theta_\pi&=&\frac{\alpha}{\sqrt{\alpha^2+\beta^2}}
\left[
1-\frac{\alpha^2\beta^2}{\left(\alpha^2+\beta^2\right)^2}
\frac{x_\pi}{y_\pi}
\right] +\cdots.
\label{sincos}
\end{eqnarray}
In these equations the first terms correspond to the
massless pion case and the second terms to the leading
$x_\pi/y_\pi$ corrections when the pion mass is turned on.

    An important application of Eq.(\ref{sincos}) is to the
axial vector current with the pion's quantum numbers.
After taking account of the mixing in
Eq.(\ref{mixingangle}), the right
hand side of the first Eq.(\ref{currents})
may be rewritten as:
\begin{equation}
(J_\mu^{axial})_1^2=F_\pi\partial_\mu\pi^+
+F_{\pi'}\partial_\mu\pi'^+ +\cdots,
\label{axcur}
\end{equation}
where,
\begin{eqnarray}
F_\pi&=&2\, \alpha \cos\theta_\pi -2\, \beta \sin\theta_\pi,
\nonumber \\
F_{\pi'}&=&2\, \alpha \sin\theta_\pi + 2\, \beta \cos \theta_\pi.
\label{dconstants}
\end{eqnarray}
In the zero pion mass case, $F_{\pi'}$ is seen to vanish.
For the non zero pion mass case, we find, using
Eq.(\ref{sincos}),
\begin{eqnarray}
F_\pi&=&2\sqrt{\alpha^2+\beta^2}+
{\cal O}
\left(
        \frac{x_\pi}{y_\pi}
\right)^2,
\nonumber \\
F_{\pi'}&=&\frac{-2\alpha^3\beta}
{\left(\alpha^2+\beta^2\right)^{3/2}}
\left(
\frac{x_\pi}{y_\pi} 
\right) +
{\cal O} 
\left(\frac{x_\pi}{y_\pi}\right)^2.
\label{2Fs}
\end{eqnarray}
It is seen that $F_\pi$ does not change much while
$F_{\pi'}$ picks up a non zero value. Thus the $\pi'$
does not decouple from the axial vector current
in the massive pion case. This means that PCAC does
not strictly hold in the massive case and hence there is
no reason to expect that the current algebra threshold theorem
should be exactly correct in the present model. 

    As a check of the accuracy of the approximate formula
(\ref{2Fs}) for $F_{\pi'}$ we made an exact numerical calculation
using the specific potential in Eq.(\ref{SpecLag}) and found
$F_{\pi'}=-6.937\times 10^{-4}$ GeV
while, for the same
parameters, the approximate formula gave
$F_{\pi'}=-6.864\times 10^{-4}$ GeV.

    Now let us discuss the pion pion scattering in the
threshold region. Our initial goal will be to see what
results may be obtained for a general choice of
chiral invariant potential, $V_0$
in Eq.(\ref{mixingLsMLag}). The general pattern of this discussion
and the notation is given in sections V, VI and VII of
\cite{bigpaper}
for the massless pion case.
 We start with the conventional $\pi-\pi$
scattering amplitude at tree level:
\begin{equation}
A(s,t,u)=-\frac{g}{2} +\sum_D
\left(
\frac{g_{8D}^2}{({\hat X}_a^2)_{DD}-s}
+\frac{g_{0D}^2}{({\hat X}_0^2)_{DD}-s}
\right). 
\label{wholeamp1}
\end{equation}
In this equation, $g$ denotes the coefficient of
the four point contact interaction among the physical (mass
diagonal) pions. Furthermore $g_{0D}$ denotes the three
point coupling constants connecting the physical pions to
the two physical
SU(3) singlet scalar mesons. Similarly $g_{8D}$ stands
for  the coupling constants connecting the physical pions to
the two physical scalar mesons which transform as the eighth
component of an SU(3) octet. The usual Mandelstam variables,
$s,t,u$ are being employed. It is straightforward to
numerically calculate
the coupling constants just mentioned and the amplitude if
the form of
$V_0$ is specified [e.g. Eq. (\ref{SpecLag})]; this will be
discussed in the next section. To proceed with the general case
we note that, for example, the coupling constant $g_{0D}$ may be
written as:
\begin{equation}
g_{0D}=
\left<
{
  {{\partial^3V}}
       \over
  {\partial{\pi}^+\partial{\pi}^- \partial(S_{0p})_D  }
}
\right> =\sum_{A,B,C}(R_\pi)_{A1}(R_\pi)_{B1} (L_0)_{CD}
\left< {{\partial^3V}\over{\partial({\phi}_1^2)_A\partial({\phi}_2^1)_B
\partial(S_0)_C}}
\right>.
\label{chainrule}
\end{equation}
As discussed in \cite{bigpaper}, there is a relation
between the three point coupling constants on the right hand side
and two point elements of the squared mass matrices. Such a
relation follows from differentiation of Eq. (\ref{geneq}) together
with the use of Eq.(\ref{mincond}). In fact it is similar in form to
the relation obtained for the zero mass pion case:
\begin{equation}
g_{0D}=\frac{2}{\sqrt{3}F_{\pi}}(R_{\pi})_{A1} (L_0)_{HD}
\left[ (X_0^2)_{AH}-(M_{\pi}^2)_{AH} \right],
\label{physicalcoupling}
\end{equation}
wherein we have now adopted the convention of
 summing over repeated indices.
The elements of the pion transformation
matrix, $(R_\pi)_{A1}$ are the angles
 $\sin \theta_\pi$ and $\cos \theta_\pi$
given in Eq. (\ref{sincos})
for the present non zero pion mass case.
In the zero pion mass case, $\sin \theta_\pi$
and $\cos \theta_\pi$ may be rewritten in terms of just $\alpha$
and $\beta$ as we see by setting the second and higher order 
terms on the right hand sides of Eq. (\ref{sincos})to zero.
That is the form in which Eq. (\ref{physicalcoupling})
 exactly holds also for the massive
pion case. Since Eq. (\ref{physicalcoupling})and its
analog for the four point vertices play an important
role in the proof of the current algebra theorem, 
we can only prove the theorem in the massive case
when we make the (not too bad) approximation that
$x_\pi/y_\pi$ is zero.  

   With this approximation we can show,
 by generalizing the treatment
given in \cite{bigpaper},
that
the usual ``current algebra" formula 
holds for the massive pion case.
 In the previous treatment, the
second term on the right hand side of Eq.(\ref{physicalcoupling})
made no contribution. Now we must take this term's contribution
into account. To get the same delicate cancellation due to
chiral symmetry we must expand the amplitude in powers of
$s-m_\pi^2$ instead of simply powers of $s$. Explicitly,
\begin{eqnarray}
A(s,t,u)&=&-\frac{g}{2} +
\left( 
\frac{g_{8D}^2}{({\hat
X}_a^2)_{DD}-({\hat M}_{\pi}^2)_{11}} +\frac{g_{0D}^2}{({\hat
X}_0^2)_{DD}-({\hat M}_{\pi}^2)_{11}}
\right)
\nonumber \\
&+& \left( s-({\hat M}_{\pi}^2)_{11} \right)
\left(
\frac{g_{8D}^2}{ \left[ ({\hat X}_a^2)_{DD}-({\hat
M}_{\pi}^2)_{11}\right]^2} +\frac{g_{0D}^2}{\left[({\hat 
X}_0^2)_{DD}-({\hat
M}_{\pi}^2)_{11}\right]^2}
\right) +\cdots.
\label{expandedamp1}
\end{eqnarray}
Note that $({\hat X}_a^2)_{DD}$ is a single number indexed by $D$.
There is a huge simplification of the coefficients of the $(s-m_\pi^2)$
term:
\begin{eqnarray}
&&\frac{g_{0D}^2}{[({\hat X}_0^2)_{DD}-({\hat M}_{\pi}^2)_{11}]^2}
\nonumber\\
&&=\frac{4}{3F_\pi^2}(R_\pi)_{A1}
\left[(X_0^2)_{AH}-(M_{\pi}^2)_{AH}\right]
(L_0)_{HD}
\frac{1}{\left[ ({\hat X}_0^2)_{DD}-({\hat
M}_{\pi}^2)_{11}\right]^2}(R_{\pi})_{C1}[(X_0^2)_{CK}-(M_{\pi}^2)_{CK}](L_0)_{KD}
\nonumber \\
&&=\frac{4}{3F_{\pi}^2}(R_{\pi}^{-1})_{1G}(L_0)_{GE}(L_0^{-1})_{EA}(X_0^2)_
{AH}
(L _0)_{HD}\frac{1}{ \left[({\hat X}_0^2)_{DD}-({\hat 
M}_{\pi}^2)_{11}\right]^2}
(L_0^{-1})_{DK}(X_0^2)_{KC}(L_0)_{CF}(L_0^{-1})_{FJ}(R_{\pi})_{J1}-
\nonumber\\
&&
\frac{8}{3F_\pi^2}
(R_\pi)_{A1}(M_{\pi}^2)_{AM}(R_{\pi})_{MP}(R_{\pi}^{-1})_{
PH}(L_0)_{HD}
\frac{1}{\left[ ({\hat X}_0^2)_{DD}-({\hat M}_{\pi}^2)_{11}\right]^2}
(L_0^{-1})_{DK}(X_0^2)_{KC}(L_0)_{CF}(L_0^{-1})_{FJ}(R_{\pi})_{J1}+
\nonumber\\
&&\frac{4}{3F_{\pi}^2}(R_{\pi})_{A1}(M_{\pi}^2)_{AM}(R_{\pi})_{MP}(R_{\pi}^{-1})
_{PH}(L_0)_{HD}
\frac{1}{\left[({\hat X}_0^2)_{DD}-({\hat M}_{\pi}^2)_{11}\right]^2}
(R_{\pi}^{-1})_{1C}(M_{\pi}^2)_{CN}(R_{pi})_{NF}(R_{\pi}^{-1})_{FK}(L_0)_{KD}
\nonumber\\
&&=\frac{4}{3F_{\pi}^2}(R_{\pi}^{-1})_{1G}(L_0)_{GE}
\frac{\left[({\hat X}_0^2)_{DD}-({\hat M}_{\pi}^2)_{11}\right]^2}
{\left[({\hat X}_0^2)_{DD}-({\hat
M}_{\pi}^2)_{11}\right]^2}
(L_0^{-1})_{EJ}(R_{\pi})_{J1}=
\frac{4}{3F_{\pi}^2}.
\label{g2m402}
\end{eqnarray}
Similarly,
\begin{equation}
\frac{g_{8D}^2}{\left[({\hat X}_a^2)_{DD}-({\hat 
M}_{\pi}^2)_{11}\right]^2}
=\frac{2}{3F_{\pi}^2}.
\label{g2m48}
\end{equation}

   Next we must consider the contribution of the terms independent
of $(s-m_\pi^2)$. The four point coupling constant $g$ is approximately 
related
to the matrix elements of the squared mass matrices (again with the
proviso that the pion transformation matrices be similarly approximated) as:
\begin{equation}
g=\frac{8}{F_{\pi}^2}
\left[
\frac{1}{3}(R_{\pi}^{-1})_{1D} \left[(
X_0^2)_{DJ}-(M_{\pi}^2)_{DJ}\right] 
(R_{\pi})_{J1}+\frac{1}{6}(R_{\pi}^{-1})_{1D}
\left[(
X_0^2)_{DJ}-(M_{\pi}^2)_{DJ}\right](R_{\pi})_{J1}
\right].
\label{gfinal1}
\end{equation}
(The analogous calculation for the zero mass pion case is
given in section VII of \cite{bigpaper}.)
Using calculations similar to Eq.(\ref{g2m402}) we also get:
\begin{eqnarray}
\frac{g_{0D}^2}{ ({\hat X}_0^2)_{DD}
-({\hat M}^2_\pi)_{11}
}=
\frac{4}{3F_{\pi}^2}(R_{\pi}^{-1})_{1D}\left[(
X_0^2)_{DJ}-(M_{\pi}^2)_{DJ}\right](R_{\pi})_{J1} \label{g2m20}
\end{eqnarray}

\begin{eqnarray}
\frac{g_{8D}^2}{({\hat X}_a^2)_{DD}-({\hat M}_{\pi}^2)_{11}}
&=&
\frac{4}{6F_{\pi}^2}(R_{\pi}^{-1})_{1D} \left[(
X_a^2)_{DJ}-(M_{\pi}^2)_{DJ}\right](R_{\pi})_{J1} \label{g2m28}
\end{eqnarray}
Putting the last three equations into Eq.(\ref{expandedamp1})
we see that
the sum of the terms independent of $(s-m_\pi^2)$
vanishes in the given approximation.
The usual formula,
\begin{equation}
A(s,t,u)=\frac{2}{F_\pi^2}(s-m_\pi^2)+\cdots,
\label{caform}
\end{equation}
is thus obtained as an approximation.

    It should be remarked that Eq. (\ref{caform})
corresponds to keeping only terms up to linear order
in the expansion for $A(s,t,u)$. That means there
are corrections, even at threshold, due to the masses
of the scalars not being infinite. To see this
and to summarize in a simple way, the preceding steps let us
expand to one higher order, adopting a condensed
notation in which $m_i$ stands for the mass of any
of the four scalars while $g_i$ stands for the
corresponding trilinear coupling constant of that
scalar with two pions. The first four terms in
 the expansion of
$A(s,t,u)$, as exactly obtained from Eq.(\ref{wholeamp1}), are
\begin{eqnarray}
A(s,t,u)&=&-\frac{g}{2}+
\sum_i\frac{g_i^2}{m_i^2-m_\pi^2}\left[
1+\frac{s-m_\pi^2}{m_i^2-m_\pi^2}+
(\frac{s-m_\pi^2}{m_i^2-m_\pi^2})^2+\cdots\right]
\nonumber \\
&\approx&(s-m_\pi^2)\left[\frac{2}{F_\pi^2}
+(s-m_\pi^2)\sum_i\frac{g_i^2}{(m_i^2-m_\pi^2)^3}+\cdots\right].
\label{correction}
\end{eqnarray}
The exact first equation contains, for each $m_i$,
a geometrical expansion in the quantity $(s-m_\pi^2)/
(m_i^2-m_\pi^2)$. Thus the radius of convergence
in s for this expression is the squared mass of the lightest
scalar singlet. To apply this expression in the 
resonance region we must, of course, unitarize the
formula in some way. Here we will be content to look
at the threshold region.
In going from the first to the second equation we used
the facts established above that
(in the approximation where $F_{\pi'}=0$):
1) the sum of the first two
terms of the first equation vanishes
and 2) the third term
of the first equation simplifies
to becomes the first,
current algebra, term
of the second equation. The third term
of the second
equation represents the leading correction
to the usual current algebra
formula. It depends on the masses of the
scalar mesons and would
vanish in a hypothetical limit (often used)
in which the scalar
meson masses are taken to infinity.
Note that every term in
the approximate amplitude vanishes
  for $s=m_\pi^2$, an unphysical point called the
Adler zero \cite{adler}.
 Our derivation shows
that the Adler zero follows
from the generating equation (\ref{geneq}),
which in turn expresses
the chiral invariance of the potential,
$V_0$ and from the saturation
of the axial vector current by the pion
field (so-called partial
conservation of the axial current).
The second equation in
(\ref{correction}) is
an approximation, though a numerically
good one, because the
saturation of the axial current has
been seen to be not
strictly accurate in the present model.

   The situation in the case of zero pion mass
\cite{bigpaper} is slightly
different. There the amplitude is
proportional to $s$
so the Adler zero occurs at $s=0$,
which is also the threshold.
Thus the
current algebra amplitude as well as the
corrections due to non-infinite
mass scalar mesons
vanish at threshold in the zero pion mass case.

    In the above we found that
the current algebra theorem for a general potential
does not seem to be exactly correct. 
This small deviation
and in addition the 
more important effect of the scalar mesons will next be
calculated exactly, by numerical means, 
for the scattering amplitude using
 the leading choice of $V_0$ discussed in section III.

\section{Pion scattering: exact numerical treatment}

    In the exact numerical treatment we do not need to
 make use of the
relations between two point and three point functions
and between
three point and four point functions since we adopt
 the specific potential $V_0$ given in Eq. (\ref{SpecLag}).
The needed quantities for calculating the scattering
amplitude are displayed in Eq. (\ref{wholeamp1}): the four
physical scalar singlet masses, the four three-point
 coupling constants conecting these scalars to two pions
and the four-point pion physical coupling constant, $g$.
These are obtained by, in turn, differentiating the potential
twice with respect to two scalar fields ($\frac{{\partial}^2}
{\partial S\partial S}$)
, two pseudoscalar fields
as well as one scalar field
 ($\frac{{\partial}^3}
{\partial S\partial \phi\partial\phi}$)                                     
 and four pseudoscalar fields
 ($\frac{{\partial}^4}
{\partial \phi\partial\phi\partial\phi\partial\phi}$).                                     
Furthermore we must use equations like Eq. (\ref{chainrule})
to relate the ``bare" amplitudes obtained by such differentiations
to the physical ones (ie, in mass diagonal bases for the fields).
The matrices transforming the fields to mass diagonal bases
are defined in Eq. (\ref{simtransf}) and are obtained by
diagonalizing the relevant squared mass matrices. 

   For our purpose we {\it define} the current algebra
 result in terms of the expansion of the tree level
amplitude $A(s,t,u)$ in powers of $(s-m_\pi^2)$, as
 displayed in Eq. (\ref{correction}).
 Specifically,
\begin{equation}
A(s,t,u)=C_0 +C_1(s-m_\pi^2)+ \cdots,
\label{Aexpansion}
\end{equation}
where,
\begin{eqnarray}
C_0&=&-\frac{g}{2}+\sum_i\frac{g_i^2}{m_i^2-m_\pi^2},
\nonumber \\                 
C_1&=&\sum_i\frac{g_i^2}{(m_i^2-m_\pi^2)^2}.
\label{C0C1}   
\end{eqnarray}
 The
current algebra result requires $C_0$ to vanish
and  
 $C_1=2/F_\pi^2$.

\begin{figure}[h]
\begin{center}
\vskip 1cm
\epsfxsize = 7.5cm
\ \epsfbox{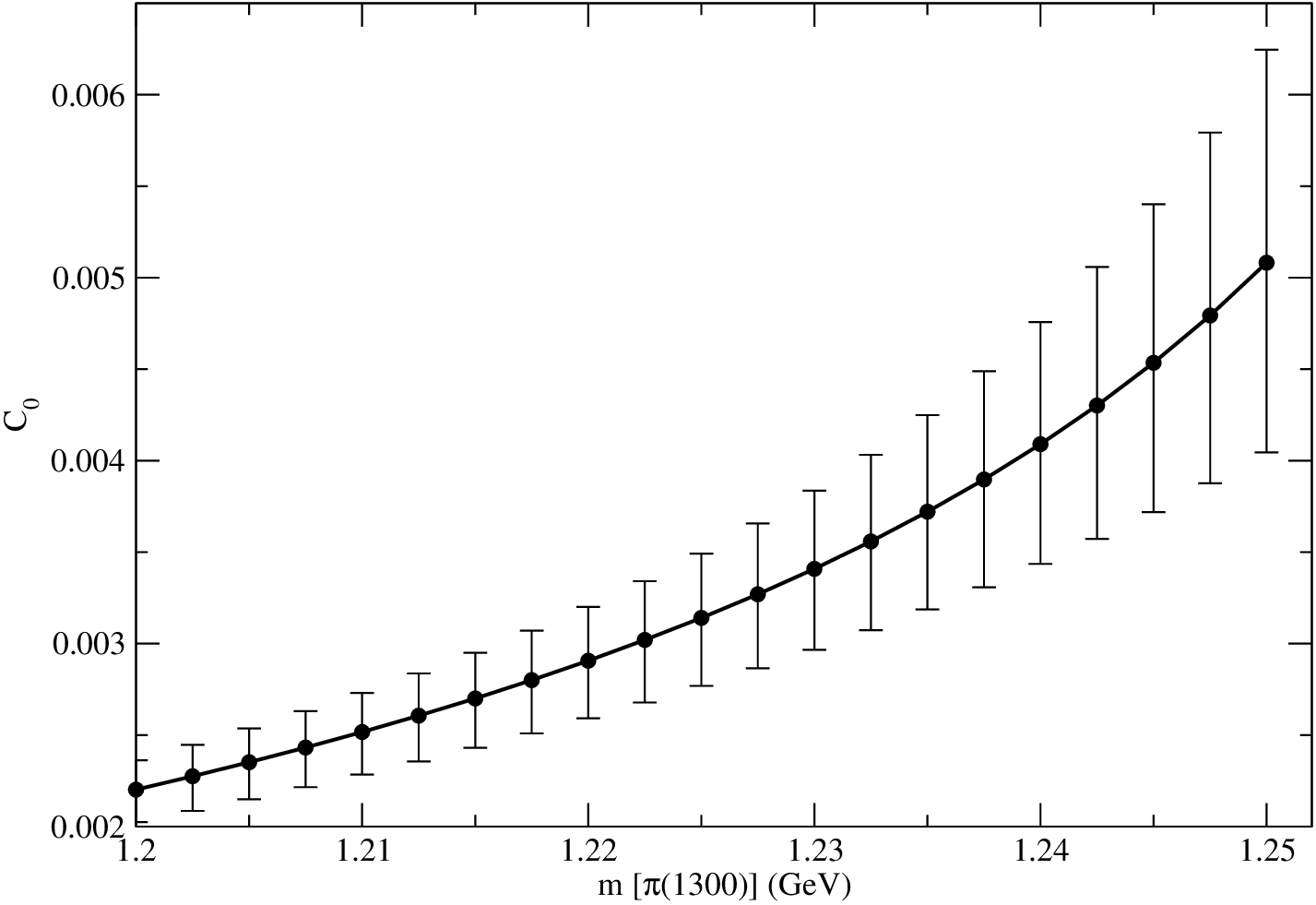}
\end{center}
\caption[]{Constant term, $C_0$ in expansion of invariant
amplitude as a function of the model parameter $m[\pi(1300)]$.
The error bars reflect the uncertainty in the mass of
$a_0(1450)$. Current algebra would predict $C_0$=0. 
}
\label{C0plot}
\end{figure}

\begin{figure}[h]
\begin{center}
\vskip 1cm
\epsfxsize = 7.5cm
\ \epsfbox{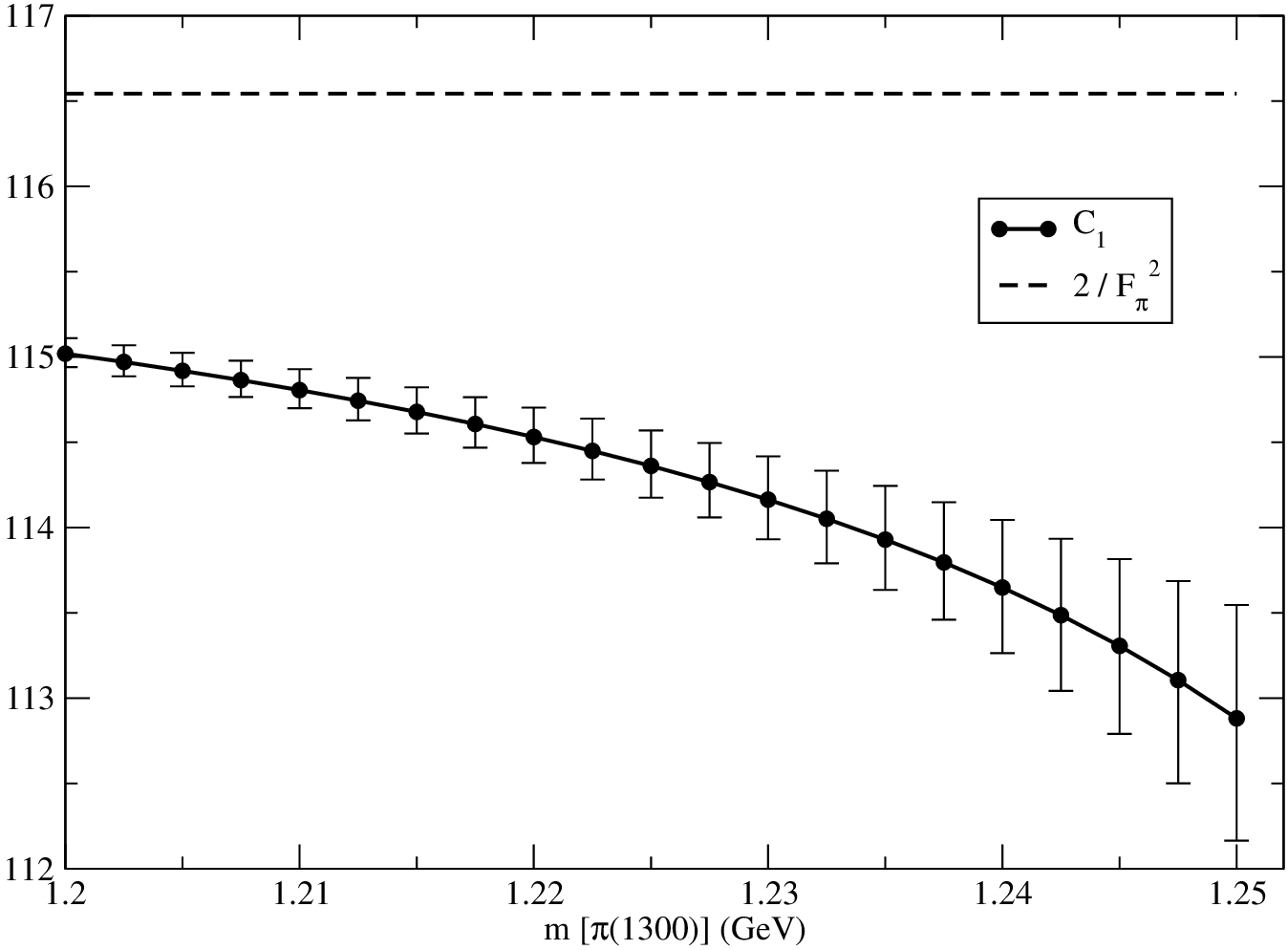}
\end{center}
\caption[]{Coefficient of linear term $C_1$
(in units of $GeV^{-2}$)
 in expansion of invariant
amplitude compared with the current algebra result
(dashed line) as a function of the 
model parameter $m[\pi(1300)]$.
The error bars reflect the uncertainty in the mass of
$a_0(1450)$. 
}
\label{C1plot}
\end{figure}

Plots of $C_0$ and $C_1$ as functions of the model
parameter $m[\pi(1300)]$ are shown in Figs. \ref{C0plot}
and \ref{C1plot} respectively. Even though $C_0$ is small
it is clearly non vanishing. Also $C_1$ deviates
by a few percent from the current algebra prediction. 
To estimate the numerical accuracy of this calculation
it was repeated for the case of zero pion mass. There it
 was found that $C_0={\cal O}(10^{-8})$ whereas
it should be exactly zero. Thus the accuracy of the
 calculation method is several orders
of magnitude more sensitive
than the indicated effect. In this model, the Adler zero
is shifted (by about -$C_0/C_1$) very slightly to
 the left of $m_\pi^2$.

   The small deviations from the current algebra result
just discussed seem to be beyond present experimental 
accuracy. On the other hand, the ``beyond" current
algebra contributions to $A(s,t,u)$ due to the higher
than linear terms in the expansion shown in
 Eq. (\ref{correction}) seem to be highly relevant
for comparison with present day experiments 
\cite{NA48,E865,dirac,an}. These corrections
would vanish in a limit where the scalar masses
all go to infinity, which essentially corresponds to the use
of a non-linear rather than the present linear type
of sigma model.

    It is usual to discuss the amplitudes near
 threshold in terms of their partial wave scattering lengths.
 The $J=0$ scattering lengths are of course especially
 affected by the presence of light scalar mesons.
Using the compact notation in Eq. (\ref{correction}),
the (dimensionless) partial wave scattering lengths
may be calculated to be: 
\begin{eqnarray}
m_\pi a_0^0&=&\frac{1}{32\pi}\left[-\frac{5g}{2}
+\sum_{i}g_i^2\left(\frac{3}{m_i^2-4m_\pi^2}
+\frac{2}{m_i^2
}\right)\right],
\nonumber \\
m_\pi a_0^2&=&\frac{1}{32\pi}\left[-g
+2\sum_{i}\frac{g_i^2}{m_i^2}\right].                  
\label{scatlen}
\end{eqnarray}
These formulas are expressed in terms of the 
physical masses and coupling constants which
 are being computed exactly by numerical means.
Note that the isospin label, $I$ and the
 angular momentum
label, $J$ appear as $a_J^I$.
    For comparison, we may give the usual current
algebra results \cite{W}:
\begin{eqnarray}
m_\pi a_0^0 &=& \frac{7m_\pi^2}{16{\pi}F_\pi^2}, 
\nonumber \\
m_\pi a_0^2 &=& \frac{-2m_\pi^2}{16{\pi}F_\pi^2}.
\label{cascatlen}
\end{eqnarray}

\begin{figure}[h]
\begin{center}
\vskip 1cm
\epsfxsize = 7.5cm
\ \epsfbox{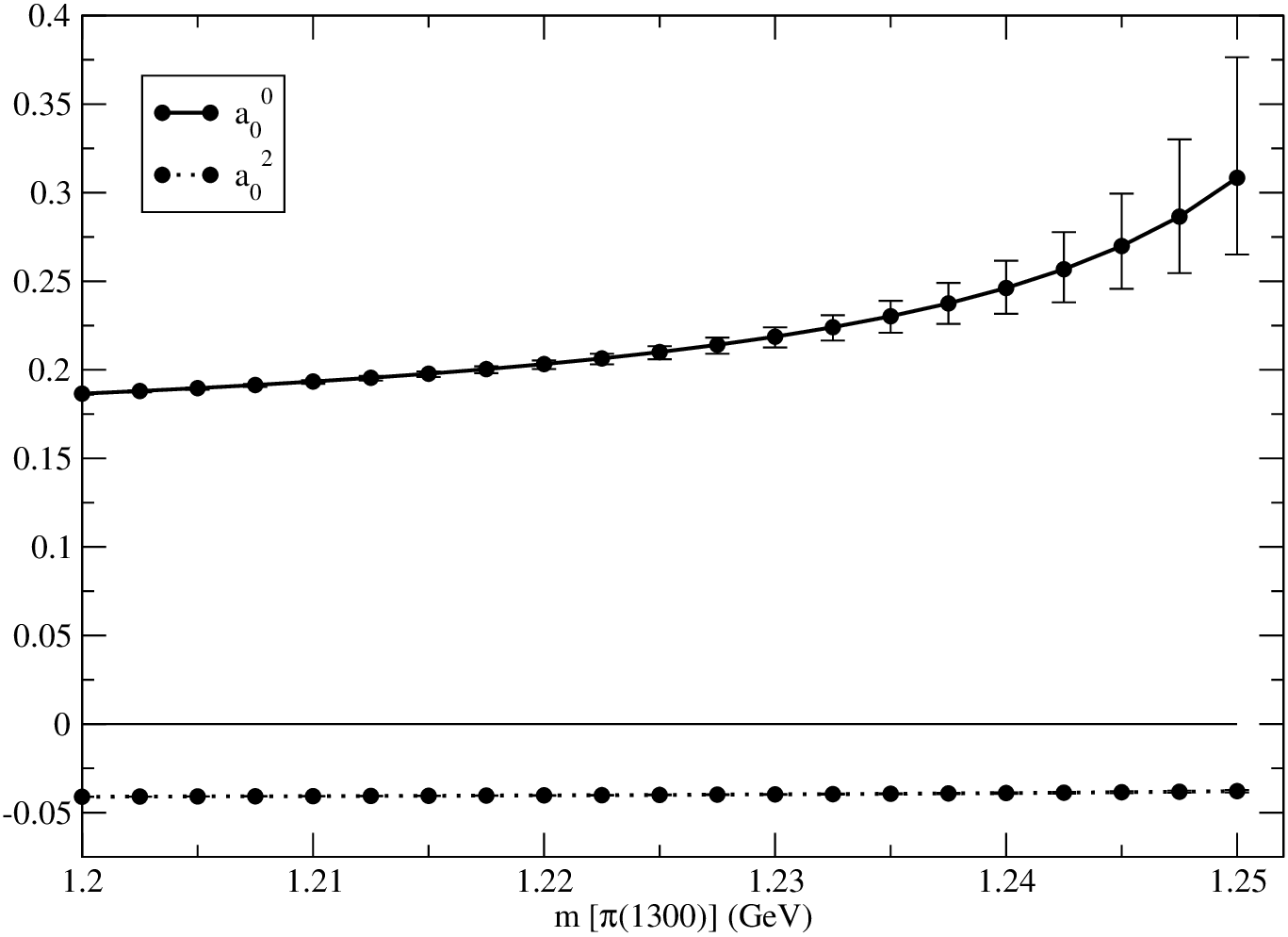}
\end{center}
\caption[]{%
Top curve: $I=J=0$ scattering length, $m_{\pi}a_0^0$ vs. $m[\pi(1300)]$.
Bottom curve: $I=2$, $J=0$ scattering length, $m_{\pi}a_0^2$ vs.
$m[\pi(1300)]$.
   The error bars reflect the
uncertainty of $m[a_0(1450)]$.
}
\label{scatle}
\end{figure}

   The results of our numerical calculation
are shown in Fig. \ref{scatle}.

It is seen that the numerical calculation for the scattering
length in the non resonant $I=2$ channel gives about the same
value, -0.04 as the current algebra result. In the 
resonant $I=0$ s-wave channel the current algebra result of
0.15 is smaller than the result of the exact calculation
for the range shown of the model parameter $m[\pi(1300)]$. 
 The exact calculation result in this resonant
channel varies strongly with 
$m[\pi(1300)]$ in contrast to the case in the
non resonant channel. To understand what this means
we should ask what is the significance of varying
$m[\pi(1300)]$. Such a variation might be
associated with variations in the masses of the four
iso-singlet scalars. But the set up of the model
as shown in Eq. (\ref{inputs1}) fixes the physical
 masses of the 
two octet isosinglets (at 985 MeV and 1474 MeV)
. Furthermore we see in Fig. 
\ref{ms0vsmpip} that varying $m[\pi(1300)]$ leaves
the mass of the heavier SU(3) singlet scalar
essentially unchanged (at about 1500 MeV)
 while it changes the 
  mass of the  lightest SU(3) singlet scalar.
Thus it seems unavoidable to interpret the variation
of $m[\pi(1300)]$ as being associated with the
 variation of the mass of the lightest scalar.
This is confirmed by noticing that the largest change
in $a_0^0$ occurs in the region of $m[\pi(1300)]$
where the mass of the lightest scalar is changing
most rapidly.

The correction to the current algebra result for 
$a_0^0$ due to the finite masses of 
light scalars was already discussed and noted to be 
positive in \cite{SU1}, some years ago. The contribution
 of a light scalar meson to the scattering length was
recently calculated in \cite{skr}.

   It is very interesting to examine the
    recent experimental data on the s-wave scattering 
lengths $a_0^0$ and $a_0^2$; these include the following.

NA48/2 collaboration \cite{NA48}:
\begin{equation}
m_{\pi^+}(a_0^0-a_0^2)=0.264 \pm 0.015
\end{equation}
\begin{equation}
m_{\pi^+}a_0^0 = 0.256 \pm 0.011
\end{equation}

E865 Collaboration \cite{E865}:
\begin{equation}
m_{\pi^+}a_0^0 = 0.216 \pm 0.015
\end{equation}

DIRAC Collaboration \cite{dirac}
\begin{equation}
m_{\pi^+}a_0^0 =0.264_{-0.020}^{+0.038}
\end{equation}

A general discussion of these experiments is given in
\cite{an}.

    Comparison of experiment with theory shows that
the larger values of $a_0^0$ predicted by the numerical
calculation when $m[\pi(1300)]$ is 
 greater than about  1215 MeV, give good agreement.
(This corresponds to the lightest scalar singlet 
lighter than about 460 MeV).
In contrast, the current algebra prediction for $a_0^0$
is clearly too low. The nonresonant channel with ($I=2$, $J=0$)
is not so well determined from experiment but seems
to be consistent with the common prediction of current
algebra or the numerical calculation.

     The indicated value of the lightest scalar mass in our model
is consistent with recent results \cite{roy} obtained by using Roy 
dispersion relation sum rules. The typical values obtained for the mass 
and width of the lightest scalar are M = 441 MeV and $\Gamma$ = 544 MeV.
These are also similar to what is obtained \cite{BFMNS01},
 M = 457 MeV and $\Gamma$
= 632  MeV, by using a
K-matrix unitarized three flavor linear sigma model (with  
just one chiral nonet). A unitarized two flavor
linear sigma model was earlier given in \cite{AS94}. The 
$s$ wave pion pion interaction has recently \cite{ay} been discussed
using the Adler sum rule. Actually, a long time ago the Adler
sum rule was used \cite{cs} to suggest a light scalar with a
 similar mass to the above.

\begin{figure}[h]
\begin{center}
\vskip 1cm
\epsfxsize = 7.5cm
\ \epsfbox{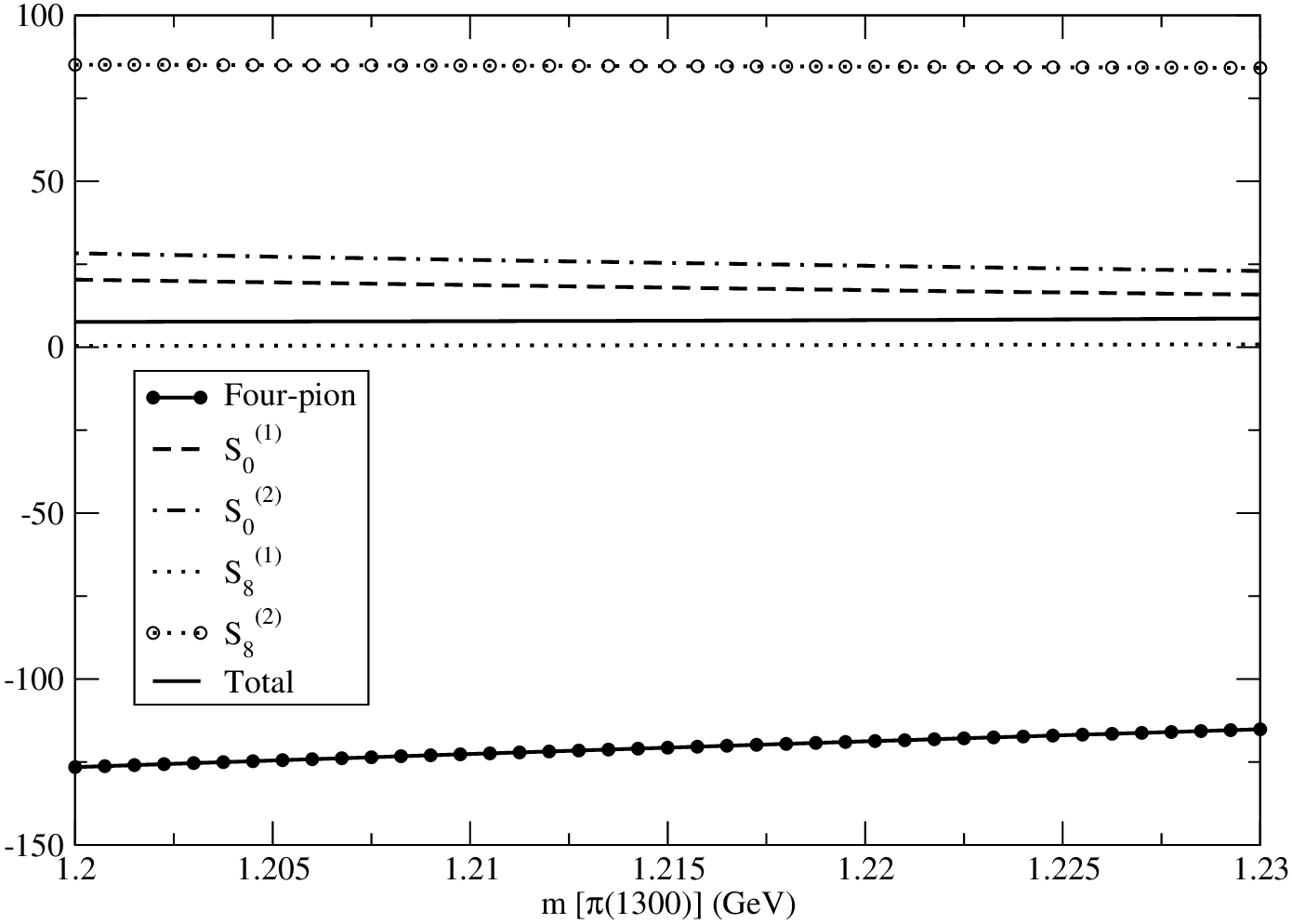}
\end{center}
\caption[]{%
 Individual contributions in Eq.(\ref{wholeamp1})
 to $A(s,t,u)$ at threshold.
}
\label{ICwholeamp1}
\end{figure}

\begin{figure}[h]
\begin{center}
\vskip 1cm
\epsfxsize = 7.5cm
\ \epsfbox{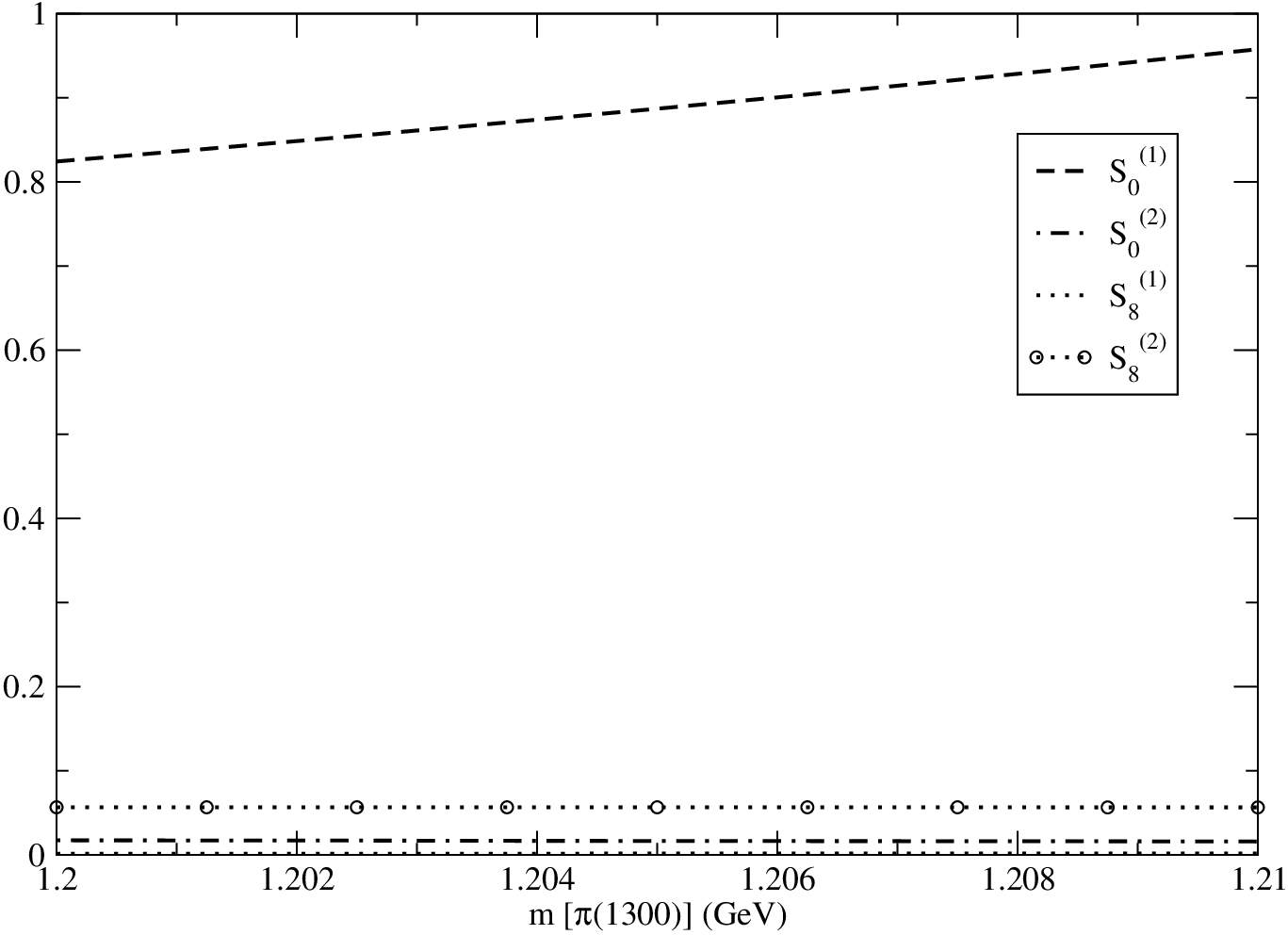}
\end{center}
\caption[]{%
Individual contributions to the
second of Eq.(\ref{correction}) to 
$A(s,t,u)$ at threshold.
}
\label{ICcorrection}
\end{figure}

    The encouraging result for $a_0^0$ with inclusion
 of scalar meson 
corrections corresponds to a tree level treatment of this
linear sigma model. One may justifiably wonder whether the
agreement would be spoiled by inclusion of loops, i.e. by a
unitarization of the model. While this is a more complicated
 matter we may note that the simplest unitarization, the 
K matrix approach, does not change the result at all. In 
this approach the corrected scattering length would be,

\begin{equation}
m_\pi a_0^0 ({\rm corrected}) = \frac{m_\pi a_0^0}{1-im_\pi a_0^0 
\sqrt{1-4m_\pi^2/s}},
\label{correctedsl}
\end{equation}
 
where the kinematical square root is to be evaluated
at threshold, $s=4m_\pi^2$. Of course this is just
an indication rather than a proof that the effect of
unitarization is expected to be small for the scattering
 lengths.

\section{Discussion and conclusions}
    
    We have seen that the intuitive explanation for the existence
of a very light scalar meson as well as light scalar mesons
with large ``four quark" content given in \cite{bigpaper}
(which used the simplifying assumption of zero pion mass)
still is good when the physical pion mass is used. The main
physical input is (almost) spontaneously broken chiral symmetry.
Our model treats the chiral symmetry in its linear realization.
This is normally considered inconvenient for the treatment
of the scattering problem since it is known that there are
very large cancellations
between the four pion contact and scalar meson pole terms
 due to the symmetry. However it has
the nice feature that it retains the light scalar mesons which give 
crucial corrections to the current algebra results. The inconvenience
due to the large cancellations can be removed by using the 
Taylor expansion in Eq. (\ref{correction}). The third
term in the second line of
 that equation shows that the beyond current algebra 
corrections can be 
neatly displayed in a physical way; they are
seen to be of order
$(m_\pi/m_i)^2$ compared to the current algebra term.
 Here $m_i$ is a scalar
meson mass.
Still higher order corrections are suppressed by
additional orders of $(m_\pi/m_i)^2$.

    It is very interesting to illustrate how the use of the 
expanded form of the amplitude given in Eq. (\ref{correction})
leads to a nice picture. In Fig. \ref{ICwholeamp1} the individual
contributions to the starting form of the amplitude,
 Eq. (\ref{wholeamp1}), associated with each scalar exchange
as well as with the four point contact term are displayed.
Evidently the contact term is the largest in
magnitude and cancels off most of the other contributions.
In addition the contribution from the
 lightest scalar $S_0^{(1)}$ exchange, 
using this starting equation, has a very small magnitude,
although its effect is expected to be the largest. Evidently,
the  intricate cancellations completely distort
 the underlying physics when expressed in this form. On the other hand,
the use of the second of Eq.(\ref{correction}) clears things up,
as may be seen from Fig. \ref{ICcorrection}. There the correction to the 
current algebra result is seen to be completely dominated by the lightest 
scalar.

   Clearly the $s$-wave pion pion scattering is rather nicely 
described by a linear sigma model. At the present stage it does not 
seem to matter which one is chosen for this aspect. The virtue 
of the present toy model
is that it accomodates both 2 quark and 4 quark scalar nonets in a consistent
way and may help in understanding the relation to QCD as earlier discussed
\cite{bigpaper}. Another interesting consequence of this linear
formulation is the presence of heavy, largely four quark
pseudoscalars, of which the $\pi(1300)$ is a possible candidate.

  An amusing feature of the present model with
a massive pion is that, due to a small deviation from exact
PCAC, the current algebra result for near threshold pion pion
 scattering no longer holds exactly. Associated with this is
the feature that the amplitude no longer vanishes at the 
unphysical point (Adler zero), $s=m_\pi^2$. The zero is shifted 
somewhat. The measure of this effect is 
the deviation of the coefficient, 
$F_{\pi'}$ from zero. We have seen that this is quite small
since $F_{\pi'}$ is of the order $x_\pi/y_\pi\approx
m_\pi^2/m_{\pi'}^2 \approx$ 0.01, as may be seen from
Eq. (\ref{2Fs}). On the other hand the similar effect
should be more noticeable for the $K-K'$ system
and for $K-{\bar K}$ scattering.  
In \cite{toyfor2} it is shown that this system is formally 
analogous to the $\pi-\pi'$ system with the substitutions
$x_\pi\rightarrow x_K$ and $y_\pi\rightarrow y_K$. 
But in this case $F_{K'}$ is expected to be of the order
 $m_K^2/m_{K'}^2\approx$ 0.1, ten times greater
than for the pion case. A full calculation would require
setting up the model with the inclusion of
SU(3) symmetry breaking quark mass terms. This will be studied
elsewhere.  

    A possible general question about the present 
model is that it introduces both states made of a quark
and an antiquark as well as states with two quarks
and two antiquarks. According to the usual 't Hooft large $N_c$
extrapolation \cite{th} of QCD the ``four quark" states
are expected to be suppressed. However, it was recently 
pointed out \cite{ss07} that the alternative, mathematically
allowed, Corrigan Ramond \cite{cr} extrapolation does not
 suppress the 
multiquark states. This kind of extrapolation may be
relevant for understanding the physics of the light scalar mesons.

\section*{Acknowledgments} \vskip -.5cm
We are happy to thank A. Abdel-Rehim, D. Black, M. Harada,
S. Moussa, S. Nasri and F. Sannino for many helpful
related discussions.
The work of A.H.F. has been partially supported by the
NSF Award 0652853.
The work of R.J. and J.S. is supported in part by the U. S. DOE under
Contract no. DE-FG-02-85ER 40231.

\appendix
\section{Parameter determination}
Given the inputs: the pion decay constant, $F_\pi$;
the mass of the pion, $m_\pi$; the mass of the
$a_0(980)$, $m_a$;
the mass of the  $a_0(1450)$, $m_{a'}$; the mass of the
$\pi(1300)$, $m_{\pi'}$, the independent model parameters
which don't involve the $U(1)_A$ violating terms can be
successively
determined (in the order given) by the equations:
\begin{eqnarray}
2 d_2 &=& {  {m_a^2 m_{a'}^2  - m_\pi^2\, m_{\pi'}^2} \over
{m_a^2 +m_{a'}^2 - m_\pi^2 - m_{\pi'}^2} }
\nonumber \\
{A\over \alpha} &=& {{m_\pi^2\, m_{\pi'}^2}\over {4\, d_2}}
\nonumber \\
(\alpha e_3^a)^2 &=&\frac{1}{64}\left( (m_a^2 - m_{a'}^2)^2  - [4 d_2 - 
(m_a^2
+m_{a'}^2)]^2\right)
\nonumber \\
4 c_2 &=&  m_a^2 + m_{a'}^2  - 2d_2 - {{ 56(\alpha e_3^a)^2} \over d_2}
- {3\over 2}\, m_\pi^2\, m_{\pi'}^2
\nonumber \\
{\beta\over \alpha} &=& {{-2 (\alpha e_3^a)} \over d_2}
\nonumber \\
{\rm cos}\, 2\theta_\pi &=&
{
   {2\, d_2 - 2\, d_2\, \left({\beta\over \alpha}\right)^2
     - 2\,\left({A\over \alpha}\right)
   }
  \over
   {
    \sqrt{
           16\, d_2^2\left(\beta\over \alpha\right)^2
          +\left[ 2\, d_2 - 2\, d_2  \left({\beta\over \alpha}\right)^2
                  -  2\,\left({A\over \alpha}\right)
           \right]^2
         }
   }
}
\nonumber \\
\alpha &=&
{1\over 2}\,
{
{F_\pi}
   \over
{ {\rm cos}\, \theta_\pi -  \left({\beta\over \alpha}\right) \,
    {\rm sin}\, \theta_\pi
}
}
\nonumber \\
c_4^a &=& {1 \over {2 \alpha^2}}
\left[ c_2 + {{8 (\alpha e_3^a)^2} \over d_2} + \left({A\over
\alpha}\right)\right]
\label{lagpara}
\end{eqnarray}

    Once the above parameters are determined, the parameters $\gamma_1$
and $c_3$ of the $U(1)_A$ violating sector are obtained in terms of
the mass of the $\eta(958)$, $m_{\eta 1}$ and the mass of a suitable
heavier
$0^-$ isosinglet, $m_{\eta 2}$ using the following procedure. The
2$\times$2 prediagonal mass-squared matrix of the two SU(3) singlet
pseudoscalars is written in the form:
\begin{equation}
(M^2_0)=
\left[ \begin{array}{c c}
                -\frac{8c_3(2\gamma_1+1)^2}{3\alpha^2} +K_{11} &
\frac{8c_3(1-\gamma_1)(2\gamma_1+1)}{3\alpha\beta}+K_{12}
\nonumber \\
      \frac{8c_3(1-\gamma_1)(2\gamma_1+1)}{3\alpha\beta}+K_{12}  &
-\frac{8c_3(1-\gamma_1)^2}{3\beta^2}+K_{22}
                \end{array} \right] ,
\label{phizeromixing}
\end{equation}
where $K_{ij}$ is a real symmetric matrix involving the
coefficients of the  terms in V$_0$
which are U(1)$_{\rm A}$ invariant . With the
choice of invariant terms in Eq.(\ref{SpecLag}) we have:
\begin{eqnarray}
K_{11} &=& - 2\, (c_2 - 2\, c_4^a\, \alpha^2 + 4\, e_3^a\, \beta)
\nonumber \\
K_{12} &=& - 8 \, e_3^a \, \alpha
\nonumber \\
K_{22} &=&  2 \, d_2
\label{K}
\end{eqnarray}

Then, $\gamma_1$ is found as a solution of the quadratic equation:
\begin{eqnarray}
0&=&S\gamma_1^2+T\gamma_1+U,
\nonumber \\
S&=&\frac{R}{\alpha^2}
\left(
       4+\frac{\alpha^2}{\beta^2}
\right)
-\frac{K_{11}}{\beta^2}
+\frac{4K_{12}}{\alpha\beta}-\frac{4K_{22}}{\alpha^2},
\nonumber \\
T&=&\frac{R}{\alpha^2}
\left(
4-2\frac{\alpha^2}{\beta^2}
\right)
+\frac{2K_{11}}{\beta^2}
-\frac{2K_{12}}{\alpha\beta}-\frac{4K_{22}}{\alpha^2},
\nonumber \\
U&=&\frac{R}{\alpha^2}
\left(
1+\frac{\alpha^2}{\beta^2}
\right)
-\frac{K_{11}}{\beta^2}
-\frac{2K_{12}}{\alpha\beta}-\frac{K_{22}}{\alpha^2},
\nonumber \\
R&=&\frac{4m_{\eta
1}^2m_{\eta 2}^2- {\rm det}  (K)}
  {m_{\eta 1}^2+ m_{\eta 2}^2-{\rm Tr} (K)}.
\label{findgamma1}
\end{eqnarray}
In addition,
\begin{equation}
c_3=\frac{\frac{3}{8}
\left(
       -m_{\eta 1}^2m_{\eta 2}^2+ {\rm det} (K)
\right)
}
{K_{11}
\left(\frac{1-\gamma_1}{\beta}
\right)^2
+2K_{12}
\left(
\frac{1-\gamma_1}{\beta}
\right)\left(\frac{1+2\gamma_1}{\alpha}\right)
+K_{22}\left(\frac{1+2\gamma_1}{\alpha}\right)^2}
\label{findc3}
\end{equation}

Next we give the numerical values of the parameters for the
central values
of all the listed input masses
except for $m[\pi(1300)]$ which instead will take the typical
value allowed by both the data and
by the model, 1215 MeV. Table \ref{T_6param} shows the results
for the parameters
which are not associated with
the  U(1)$_{\rm A}$ violating part of the Lagrangian.

\begin{table}[htbp]
\begin{center}
\begin{tabular}{c|c}
\hline \hline
$c_2 ({\rm GeV}^2)$    & 8.79 $\times 10^{-2}$ \\
$d_2  ({\rm GeV}^2)$   & 6.30 $\times 10^{-1}$\\
$e_3^a  ({\rm GeV})$   & $-2.13$\\
$c_4^a $               & 42.4  \\
$\alpha  ({\rm GeV})$  & 6.06 $\times 10^{-2}$\\
$\beta  ({\rm GeV})$   & 2.49 $\times 10^{-2}$\\
$ A ({\rm GeV}^3)$   & 6.66 $\times 10^{-4}$
\\
\hline
\end{tabular}
\end{center}
\caption[]{
Calculated Lagrangian parameters:$c_2$, $d_2$, $e_3^a$, $c_4^a$
and vacuum values: $\alpha$, $\beta$.
}
\label{T_6param}
\end{table}

Table \ref{T_2param} shows the calculated Lagrangian
parameters associated with the U(1)$_{\rm A}$ violating terms.
Two ``scenarios" associated with different identifications
of the heavy $\eta$ which is the partner of the $\eta(958)$
are shown (I assumes $\eta(1475)$ to be chosen while II
assumes $\eta(1760)$ to be chosen.) For each scenario, the
two solutions (labeled 1 and 2) are shown.

\begin{table}[htbp]
\begin{center}
\begin{tabular}{c|c|c|c|c}
\hline \hline
                   &       I1             & I2
                   &       II1            & II2\\
$c_3 ({\rm GeV}^4)$&$-2.39 \times 10^{-4}$&$-2.38 \times 10^{-4}$
                   &$-3.42 \times 10^{-4}$& $-3.37 \times 10^{-4}$\\
$\gamma_1$         & 5.33 $\times 10^{-1}$ &  $2.52 \times 10^{-1}$
                   & 8.68 $\times 10^{-1}$& $-8.65 \times 10^{-2}$\\
\hline
\end{tabular}
\end{center}
\caption[]{
Calculated parameters: $c_3$ and $\gamma_1$.
}
\label{T_2param}
\end{table}

Using these parameters we next list the mixing matrices for,
respectively, the two $0^-$ octet states, the two $0^+$ octet
states and the two $0^+$ singlet states:

\begin{equation}
(R_\pi^{-1})  =
\left[
\begin{array}{cc}
0.923 & $0.385$ \\
-$0.385$ & 0.923\\
\end{array}
\right], \hspace{.3cm}
(L_a^{-1})  =
\left[
\begin{array}{cc}
-$0.486$ & $0.874$ \\
0.874    & 0.486\\
\end{array}
\right], \hspace{.3cm}
(L_0^{-1})  =
\left[
\begin{array}{cc}
0.706 & 0.708 \\
$0.708$    & -0.706\\
\end{array}
\right].
\label{mms}
\end{equation}

Similarly, the mixing matrices for the two solutions for
scenario I of the $0^-$ singlet states are:

\begin{equation}
I\,\,1:(R_0^{-1})  =
\left[
\begin{array}{cc}
-$0.671$ & 0.742 \\
0.742 & 0.671\\
\end{array}
\right],     \hspace{.3cm}
I\,\,2: (R_0^{-1})=
\left[
\begin{array}{cc}
0.853 & -$0.522$ \\
0.522 & 0.853\\
\end{array}
\right].
\label{R0_num_I}
\end{equation}

Finally, the mixing matrices for the two solutions for
scenario II of the $0^-$ singlet states are:

\begin{equation}
II\,\,1:(R_0^{-1})  =
\left[
\begin{array}{cc}
-$0.411$ & 0.912 \\
0.912 & 0.411\\
\end{array}
\right], \hspace{.3cm}
II\,\,2: (R_0^{-1})  =
\left[
\begin{array}{cc}
0.972 & -$0.235$ \\
0.235 & 0.972\\
\end{array}
\right].
\label{R0_num_II}
\end{equation}

\end{document}